\documentclass[preprint,12pt,3p]{elsarticle}

\usepackage{graphicx}
\usepackage{epsfig}
\usepackage{amssymb} \usepackage{amsthm}
\usepackage{amsfonts} \usepackage{amsmath}
\usepackage[latin5]{inputenc}

\biboptions{sort&compress}

\journal{Chinese Journal of Physics}

\begin{document}

\begin{frontmatter}

\title{Action in Hamiltonian Models Constructed by Yang-Baxter Equation: Entanglement and Measures of Correlation}

\author[label1]{Durgun Duran\corref{cor1}}
\address[label1]{Department of Physics, Yozgat Bozok University, Faculty of Science
and Arts, 66100, Yozgat, Turkey.}

\cortext[cor1]{Corresponding author.}

\ead{durgun.duran@bozok.edu.tr}


\begin{abstract}
By using the quantum Yang-Baxterization approach, we investigate the dynamics of quantum entanglement under the actions of different Hamiltonians on the different two-qubit input states and analyze the
effects of the Yang-Baxter operations on it. During any quantum process that takes place in a noisy environment, quantum correlations display behavior that does not increase. We point out that for
two-qubit systems subject to actions of different Yang-Baxter operations the loss of correlations can be mitigated by the appropriate choice of the initial states and the Yang-Baxterization process.
We show that in a noisy environment it possible to create the optimal conditions for performing any quantum information task.
\end{abstract}


\end{frontmatter}


\section{Introduction}
\label{sec1}

Quantum correlations are the milestone resources for performing various quantum information and computation tasks, such as key distribution, cryptography, superdense coding and teleportation
which are not classically plausible \cite{NC}. In carrying out such a task, preservation and maintaining of correlations for a long duration have vital importance \cite{DiVincenzo}.
However, it is well-known that they decrease under any quantum operation such as quantum channels in a noisy environment \cite{PlenioVirmani}. In fact, loss of correlations called decoherence
is the main snag in real-world applications of new technologies based on quantum information and computational sciences \cite{Breuer,Schlosshauer2008}.
Therefore, searching for new ways of controlling the decrease of correlations and making them available in information technologies have a great deal of interest \cite{Schlosshauer2008,china}.

The well-known measure of the quantum correlations living in a bipartite system that we shall deal with is the entanglement of formation (EoF) which quantifies the minimal cost needed to
prepare a certain quantum state in terms of maximally entangled pairs and the required amount of quantum communication \cite{Bennett, Wootters1998, Werner, Terhal2000, Rungta2003}.

Another measure of quantum correlations is the quantum coherence that originates from the description of the wave function of quantum systems and cannot be described by the classical physics laws.
It can be said that there are quantum states which have no classical analog as a result of quantum coherence and this can only be expressed in character by the laws of quantum mechanics \cite{Glauber,Sudarshan}.
These states play an essential role in the achievement of quantum supremacy \cite{Harrow}. In fact, quantum consistency is widely accepted as a key resource in the context of quantum information
processing \cite{Streltsov,Rana,Hu2017}, and thus it is very important to quantify the amount of coherence present in a quantum state. Coherence is very fragile and inevitably tends to environmental effects
due to realistic systems that interact with their external environment. This clearly means that quantum coherence is usually very difficult to be created, sustained and manipulated in
quantum systems. Therefore, it is very crucial and remarkable to create, maintain and preserve the quantum coherence in quantum computation and quantum information processing.

The idea of quantum coherence has a great interest in various other physical fields such as the extraction of work from quantum coherence in the field of quantum thermodynamics \cite{Korzekwa},
low-temperature thermodynamics with quantum coherence \cite{Narasimhachar,Lostaglio}, the role in energy transfer \cite{Henao}, quantum biology \cite{Lloyd}, photosynthesis \cite{Lambert},
the avian compass in migratory birds \cite{Gauger}, quantum algorithms \cite{Chuang,Gershenfeld} and quantum metrology \cite{Joo,Giovannetti}.

The last measure of quantum correlation is the measurement-induced disturbance (MID) that draws on the fact that measurements of quantum systems disturb them to capture quantum correlations
\cite{Luo2008}. MID can be considered as quantum discord that involves finding a local projector (on subsystem B or A) that minimizes its value. Moreover, discord is not symmetric, meaning that the
discord of A with respect to B is not the same as that of B with respect to A \cite{Zurek,Henderson}. That is why one has to think which subsystem is more appropriate to perform the local measurement
on to define discord and it measures the nonclassicality of bipartite correlations via the difference between quantum versions of two classically equivalent expressions for mutual information
\cite{Vedral2010}. However, there are two differences between them. First, the local measurements are performed on both subsystems A and B. Additionally, MID does not require searching for the optimal
set of local projectors. Instead, the chosen projectors are constructed using the eigenvectors of the reduced density matrices of A and B.

Recently, the Yang-Baxter equations (YBEs) have been introduced to the field of quantum information and quantum computation. It has been shown that YBE has a notable connection with entanglement swapping
and topological quantum computation \cite{Kitaev,Kauffman,Franko,Zhang1,Zhang2,Zhang3,Chen1,Chen2,Chen3}. The universal quantum gates usually identify the unitary solution of the braided Yang-Baxter
(the braid group relation) and unitary solutions of the quantum YBE \cite{Brylinski,Wang}. This provides a novel and significant way to examine the quantum entanglement via quantum YBEs. Generally,
a Hamiltonian can be constructed from the unitary $R(\theta, \varphi)$ matrix by the Yang-Baxterization approach. In recent years, for the unitary evolution of entangled states, this approach has been
exploited to derive a Hamiltonian \cite{Jones,Ge1991}. It has been shown that YBE can also be tested for various quantum optics applications. \cite{Hu2008}. In another study, it is found that any pure
two-qudit entangled state can be achieved by a universal Yang-Baxter matrix (YBM) which satisfies the (universal) quantum Yang-Baxter equation assisted by local unitary transformations \cite{Chen3}.
In recent work, the sudden death of entanglement has been investigated in constructed Yang-Baxter systems (YBSs) \cite{Hu2010} which are the various extensions of the YBEs for several matrices\cite{Friedel,Nijhoff,Hlavaty}.

Differently from previous works we shall study the behavior of quantum correlations and advantages and deficiencies of these measures compared to each other under the three Hamiltonians constructed by YBE.
We use EoF, $l_1$-norm of coherence and MID as measures of correlations of a two-qubit system initially prepared in parameterized Werner-like states. Then we study the actions of YBEs on these measures.
In general, although quantum correlations monotonically decrease under quantum operations, we observe that relative increments are possible to enhance the quantum correlations with an appropriate choice
of parameters and the help of YBEs.

This study is structured as follows. In Sec. 2 the main traits of EoF, quantum coherence and MID that will be used in due course are summarized. The YBM which satisfy the (universal) quantum Yang-Baxter
equation and their properties are carried out in Sec 3. Hamiltonian models that will be investigated are considered in Sec. 4 where the actions of YBM on the Hamiltonians are given. The main results of
this work are emphasized in Sec 5. We end up with some concluding remarks.

\section{Quantum Correlation Measures}
In this section, we aim to briefly describe these three measures along with some other proposed quantifiers to compare them and find out their interrelations and limitations.
To begin with, we describe the EoF, quantum coherence (relative entropy of coherence and $l_1$-norm of coherence) and MID in three sections.
\subsection{Entanglement of formation}
Firstly, as a measure of quantum correlations EoF is monotonically increasing function of the so-called concurrence $C(\rho_{AB})$ which can be defined as \cite{Wootters1998},
\begin{eqnarray}
C(\rho_{AB})=\max \{0, \lambda_1-\lambda_2-\lambda_3-\lambda_4\},
\end{eqnarray}
for any two-qubit state $\rho_{AB}$. Here the $\lambda_i$'s are the square roots of the eigenvalues of non-Hermitian matrix $\rho_{AB}\tilde\rho_{AB}$ in descending order
$\lambda_1\geq\lambda_2\geq\lambda_3\geq\lambda_4$ and $\tilde\rho_{AB}=(\sigma_y\otimes\sigma_y)\rho_{AB}^*(\sigma_y\otimes\sigma_y)$ is the time-reversed (or the spin-flipped)
state corresponding to $\rho_{AB}$. $\sigma_y$ is the Pauli $y$-matrix and asteriks denotes the complex conjugation in the two-qubit computational basis. Particularly,
for any pure state $\rho_{AB}=|\psi\rangle\langle \psi|$ where $|\tilde\psi\rangle=(\sigma_y\otimes\sigma_y)|\psi^*\rangle$, $C(\rho_{AB})=|\langle \psi|\tilde\psi\rangle|$
\footnote{For any pure two-qubit state $\rho=|\psi\rangle\langle \psi|$ the matrix $T=\rho\tilde\rho$ is of the form
$T=\langle \tilde\psi|\psi\rangle|\psi\rangle\langle \tilde\psi|$ and obeys the equation $T(T-|\langle \psi|\tilde\psi\rangle|^2)=0$.
This shows that the only nonzero eigenvalue of $T$ is $|\langle \psi|\tilde\psi\rangle|^2$ and therefore $C(\rho)=|\langle \psi|\tilde\psi\rangle|$.}.

For X-type states (such as Werner-like, isotropic and Bell-diagonal states) Eq. (1) reduces to
\begin{eqnarray}
C(\rho_{AB})=2\max \{0, |\rho_{23}|-\sqrt{\rho_{11}\rho_{44}},
|\rho_{14}|-\sqrt{\rho_{22}\rho_{33}}\}
\end{eqnarray}
where $\rho_{ij}$ are the matrix elements of $\rho_{AB}$  \cite{Eberly}.

In terms of the binary entropy
\begin{eqnarray}
h(x)=-x\log_2x-(1-x)\log_2(1-x),
\end{eqnarray}
and the concurrence $C=C(\rho_{AB})$ we have an analytic formula
\begin{eqnarray}
E(C)=h\Big(\frac{1+\sqrt{1-C^2}}{2}\Big),
\end{eqnarray}
for EoF that is a monotonically increasing function of $C$. $C=0$ ($E(0)=0$) and $C=1$ ($E(1)=1$) correspond to an unentangled state and a maximally entangled state, respectively.

\subsection{Quantum coherence}
Secondly, the relative entropy of coherence present in a quantum state represented by the two-qubit density matrix $\rho_{AB}$ or shortly $\rho$ is defined as \cite{Baumgratz}
\begin{eqnarray}
C_r(\rho)=S(\rho_{diag})-S(\rho),
\end{eqnarray}
where $S(\rho)$ is the von Neumann entropy of $\rho$ and $\rho_{diag}$ denotes the state obtained from $\rho$ by removing all the off-diagonal elements of $\rho$. Note that $C_r$ is a
basis-dependent quantity. Formally, due to its similarity to that of the relative entropy of entanglement, $C_r$ specifically represents the optimal rate of the distilled maximally
coherent states. These states can be produced by incoherent operations in the asymptotic limit of many copies of density matrix $\rho$ \cite{Winter}. Interestingly, the experimental measurement
of this coherence quantifier is also possible without full quantum state tomography \cite{Yu2017}.

The $l_1$ norm of coherence in which we focus on in this paper is given by \cite{Baumgratz}
\begin{eqnarray}
C_{l_1}(\rho)=\sum_{i\neq j}|\rho_{ij}|,
\end{eqnarray}
This measure of coherence which is basis-dependent like $C_r$ is currently not known to have any analog in the entanglement resource theory \cite{Streltsov}.

Interestingly, for any $d$-dimensional mixed state it has been proved that $C_{l_1}(\rho)\geq C_r(\rho)/\log_2 d$ and conjectured that $C_{l_1}(\rho)\geq C_r(\rho)$ for all states \cite{Rana}.

\subsection{Measurement-induced disturbance}
Finally, we introduce the MID as a measure of correlations. It is defined in terms of the quantum mutual information which quantifies the total (classical and quantum)
correlations living in a bipartite system as follows \cite{Luo2008,Datta}
\begin{eqnarray}
MID(\rho)=I(\rho)-I[\Pi(\rho)],
\end{eqnarray}
where
\begin{eqnarray}
\Pi(\rho)=\sum_{i=1}^k \sum_{j=1}^l (\Pi_i^{A}\otimes \Pi_j^{B})\rho(\Pi_i^{A}\otimes \Pi_j^{B}).
\end{eqnarray}
is the classical state closest to the initial state $\rho$ since such a measurement can leave the reduced states invariant.
Here, $\Pi_i^{A}$ and $\Pi_j^{B}$ are projectors constructed using the eigenvectors of the reduced density matrices of systems A and B, respectively. Note that for two-qubit systems,
and hence the X states we focus on in this paper, $k, l=2$. In fact, for X states, $\Pi(\rho)$ is written as $\Pi_X(\rho)=diag(\rho_{11},\rho_{22},\rho_{33},\rho_{44})$ in the two-qubit
computational basis. So, MID is given by
\begin{eqnarray}
MID_X(\rho)=-S(\rho_X)+S[\Pi_X(\rho)],
\end{eqnarray}
where $S(\rho_X)$ is the von Neumann entropy of the X-state before measurement and $S[\Pi_X(\rho)]$ is the von Neumann entropy of the density matrix after the
measurement. A plausible distance between $\rho$ and $\Pi(\rho)$ can be used to measure the quantum correlations in $\rho$.

\section{Yang-Baxter Matrices}
The representations of the Artin braid group and more specifically by using solutions to the YBE \cite{Yang,Baxter} that first discovered concerning $1+1$ dimensional quantum field theory and
two-dimensional models in statistical mechanics can construct a class of invariants of knots and links called quantum invariants. Braiding operators play a crucial role in constructing representations
of the Artin braid group, and in the construction of invariants of links and knots. The association of a Yang-Baxter operator $R$ to each elementary crossing in a link diagram is an important concept
in the construction of quantum link invariants.

The operator $R$ is a linear mapping \cite{Kauffman} $R: V\otimes V \rightarrow V\otimes V$ defined on the two-fold tensor product of a vector space $V$ generalizing the permutation of the factors
(i.e., generalizing a swap gate when $V$ represents one qubit). Such transformations need not be unitary in topological applications. The unitary $R$-matrices can be used to construct
unitary representations of the Artin braid group.

A solution to the YBE, as described above is a matrix $R$, regarded as a mapping of a two-fold tensor product of a vector space $V\otimes V$ to itself that satisfies the equation
\begin{eqnarray}
(R\otimes \mathbb{I})(\mathbb{I}\otimes R)(R\otimes \mathbb{I})=(\mathbb{I}\otimes R)(R\otimes \mathbb{I})(\mathbb{I}\otimes R),
\end{eqnarray}
where $\mathbb{I}$ is the identity operator.

In this paper, we need to study solutions of the YBE that are unitary to relate quantum computing and quantum entanglement. Then the $R$ matrix can be seen either as a braiding matrix or as
a quantum gate in a quantum computer.

The unitary $R$-matrix satisfies the YBE
\begin{eqnarray}
R_i(\mu)R_{i+1}(\mu+\nu)R_i(\nu)=R_{i+1}(\nu)R_i(\mu+\nu)R_{i+1}(\mu),
\end{eqnarray}
or
\begin{eqnarray}
R_{i}(\mu)R_{i+1}\left(\frac{\mu+\nu}{1+\beta^2\mu\nu}\right)R_{i}(\nu)=R_{i+1}(\nu)R_{i}\left(\frac{\mu+\nu}{1+\beta^2\mu\nu}\right)R_{i+1}(\mu),
\end{eqnarray}
where $\beta =-i/c$ (c is the velocity of light)\cite{Jimbo1989}, $\mu$ and $\nu$  are the parameters which usually range over the real numbers $\mathbb{R}$ in the case of an additive parameter,
or over positive real numbers $\mathbb{R^+}$ in the case of a multiplicative parameter. It is worth noting that the four-dimensional YBE Eqs. (11) and (12) admit the Temperly-Lieb algebra (TLA) \cite{Temperley,Hu2007}.
Actually the rational solution of the YBE, $R(\mu)$ can be written in terms of a unitary transformation $U$ in the following way: $R(\mu) = a(\mu)\mathbb{I} + b(\mu)U$, with $U$ satisfying the TLA
\begin{eqnarray}
U_iU_{i+1}U_i=U_i,\quad U_{i}^2=dU_i,\quad U_iU_{j}=U_{j}U_i
\end{eqnarray}
for $|i-j|\geq2$, where $d$ is the single loop in the knot theory which does not depend on the sits of the lattices. When $d=2$, the Hermitian matrix $U$ has forms as follows
\begin{eqnarray}
U_{1}=\left( {\begin{array}{cccc}
1 & 0 & 0 & e^{i\varphi}\\
0 & 0 & 0 & 0\\
0 & 0 & 0 & 0\\
e^{-i\varphi} & 0 & 0 & 1
\end{array}}
\right),\quad
U_{2}=\left( {\begin{array}{cccc}
0 & 0 & 0 & 0\\
0 & 1 & e^{i\varphi} & 0\\
0 & e^{-i\varphi} & 1 & 0\\
0 & 0 & 0 & 0
\end{array}}
\right),
\end{eqnarray}
When $d=\sqrt{2}$, the Hermitian matrix U takes the form
\begin{eqnarray}
U_{3}=\frac{1}{\sqrt{2}}\left( {\begin{array}{cccc}
1 & 0 & 0 & e^{i\varphi}\\
0 & 1 & i\varepsilon & 0\\
0 & -i\varepsilon & 1 & 0\\
e^{-i\varphi} & 0 & 0 & 1
\end{array}}
\right)
\end{eqnarray}
where $\varphi$ is real and $\varepsilon=\pm$.

Three unitary matrices $R_{i,i+1}(\theta, \varphi)$ are obtained by the Yang-Baxterization approach \cite{Jones,Ge1991} according to the above $U$ matrices as follows,
\begin{subequations}
\begin{align}
R_{i,i+1}^{(1)}(\theta, \varphi)=&\left(\cos\frac{\theta}{2}+\frac{i}{2}\sin\frac{\theta}{2}\right)\mathbb{I}_i \mathbb{I}_{i+1}-2i\sin\frac{\theta}{2} S_i^z S_{i+1}^z \nonumber\\
&-i\sin\frac{\theta}{2}\left(e^{i\varphi}S_i^+ S_{i+1}^{+}+e^{-i\varphi}S_i^- S_{i+1}^-\right),\\
R_{i,i+1}^{(2)}(\theta, \varphi)=&\left(\cos\frac{\theta}{2}+\frac{i}{2}\sin\frac{\theta}{2}\right)\mathbb{I}_i \mathbb{I}_{i+1}+2i\sin\frac{\theta}{2} S_i^z S_{i+1}^z\nonumber\\
&-i\sin\frac{\theta}{2}\left(e^{i\varphi}S_i^+ S_{i+1}^{-}+e^{-i\varphi}S_i^- S_{i+1}^+\right),\\
R_{i,i+1}^{(3)}(\theta, \varphi)=&-\cos\frac{\theta}{2}\mathbb{I}_i \mathbb{I}_{i+1}-i\sin\frac{\theta}{2}(e^{i\varphi}S_i^+ S_{i+1}^{+}+e^{-i\varphi} S_i^{-} S_{i+1}^{-})\nonumber\\
&+\varepsilon\sin\frac{\theta}{2}(S_i^+ S_{i+1}^{-}-S_i^- S_{i+1}^+),
\end{align}
\end{subequations}
where $S_i^z$ is the spin operators for the ith particle and $S_i^{\pm} = S_i^{x} \pm iS_i^{y}$ are raising and lowering operators respectively for the ith particle. The parameter $\theta$ appearing
in Eqs. (16a) and (16b) is related to $\mu$ as $\cos\theta=(1-\mu^2)/(1+\mu^2)$. In Eq. (16c), the relation of $\theta$ and $\mu$ can be written as $\cos\theta=1/\cosh\mu$. Note that solutions of the YBE
for $d=2$ are given by meromorphic functions of $\mu$ whereas for $d\neq2$ by trigonometric functions. The difference of two $\theta$ and $\mu$ relations come from this property of YBE.

\section{Dynamical Models}
Consider a system of two spin-$1/2$ particles (particle $1$ and $2$) or nearest spin-spin interaction described by an initial Hamiltonian $H_0$ \cite{Hu2010,Sun2009}
\begin{eqnarray}
H_0=\mu_1 S_1^z+\mu_2 S_2^z+gS_1^zS_2^z,
\end{eqnarray}
where $\mu_i$ represents external magnetic field and $g$ is the coupling constant of z-component of two neighboring spins. For convenience of calculations,
we introduce two parameters $B=(\mu_1+\mu_2)/2$ and $J=(\mu_1-\mu_2)/2$. Taking into account the Schr\"{o}dinger equation
\begin{eqnarray}
i\hbar\frac{\partial}{\partial t}|\Psi(\theta,\varphi)\rangle=H(\theta,\varphi)|\Psi(\theta,\varphi)\rangle
\end{eqnarray}
and $|\Psi(\theta,\varphi)\rangle=R(\theta, \varphi)|\Psi_0\rangle$ one obtains
\begin{eqnarray}
i\hbar\frac{\partial}{\partial t}R(\theta, \varphi)|\Psi_0\rangle=H(\theta,\varphi)R(\theta, \varphi)|\Psi_0\rangle
\end{eqnarray}
where $|\Psi_0\rangle$ is the eigenstate of $H_0$. Let real parameters $\theta$ and $\varphi$ be time-independent, one can get a Hamiltonian through the unitary transformation
$R(\theta, \varphi)$ as $H(\theta,\varphi)=R(\theta, \varphi)H_0 R^{-1}(\theta, \varphi)$. Now, three Hamiltonians are obtained from Eqs. (16) as follows \cite{Hu2010}
\begin{subequations}
\begin{align}
    H_1(\theta, \varphi)=&B\cos\theta(S_1^z+S_2^z)+J(S_1^z-S_2^z)+gS_1^zS_2^z\nonumber\\
    &+iB\sin\theta\left(e^{i\varphi}S_1^+ S_2^{+}-e^{-i\varphi}S_1^- S_2^-\right),\\
    H_2(\theta, \varphi)=&B(S_1^z+S_2^z)+J\cos\theta(S_1^z-S_2^z)+gS_1^zS_2^z\nonumber\\
    &+iJ\sin\theta\left(e^{i\varphi}S_1^+ S_2^{-}-e^{-i\varphi}S_1^- S_2^+\right),\\
    H_3(\theta, \varphi)=&B\cos\theta(S_1^z+S_2^z)-iB\sin\theta\left(e^{i\varphi}S_1^+ S_2^{+}-e^{-i\varphi}S_1^{-} S_2^{-}\right)\nonumber\\
    &+gS_1^zS_2^z+J\cos\theta(S_1^z-S_2^z)+\varepsilon J\sin\theta\left(S_1^+ S_2^{-}+S_1^- S_2^+\right).
\end{align}
\end{subequations}

Specifically, for $\varphi=-\pi/2$ and 2-qubits we can see that the second model and the third model are respectively the anisotropic Heisenberg XXZ chain and anisotropic Heisenberg XYZ chain under an
inhomogeneous magnetic field.

\section{Dynamics of entanglement in Yang-Baxter Systems}
In this section, we investigate the dynamics of entanglement using EoF, quantum coherence and MID as measures of correlations for three Hamiltonians under the adjoint action of unitary YBE $R(\theta, \varphi)$
on the 2 parameterized bipartite two-qubit input states which have the form Werner-like states.

For convenience, we set $\theta=\pi/2$, henceforward. So, we can rewrite Hamiltonians in Eqs. (20) as follows ($\varepsilon=1$)
\begin{subequations}
\begin{align}
H_1( \varphi)=&J(S_1^z-S_2^z)+gS_1^zS_2^z+iB\left(e^{i\varphi}S_1^+ S_2^{+}-e^{-i\varphi}S_1^- S_2^-\right),\\
H_2(\varphi)=&B(S_1^z+S_2^z)+gS_1^zS_2^z+iJ\left(e^{i\varphi}S_1^+ S_2^{-}-e^{-i\varphi}S_1^- S_2^+\right),\\
H_3(\varphi)=&J\left(S_1^+ S_2^{-}+S_1^- S_2^+\right)+gS_1^zS_2^z-iB\left(e^{i\varphi}S_1^+ S_2^{+}-e^{-i\varphi}S_1^{-} S_2^{-}\right).
\end{align}
\end{subequations}

\subsection{Action of $H_1$ to Initial Werner-like States and Entanglement Dynamics}

For a general initial input states $\rho$, the output $\sigma$ under the unitary time evolution or unitary adjoint action $ad_{U}(\cdot)=U(\cdot)U^{\dagger}$ is found to be
\begin{eqnarray}
\sigma=U\rho U^{\dagger}=e^{-itH}\rho e^{itH}.
\end{eqnarray}

Firstly, we investigate the dynamics of entanglement for the outputs under action of Hamiltonian $H_1$ on two initial Werner-like states. We first fix the two-qubit state (Werner state) to be
\begin{eqnarray}
\rho_{AB}^{(1)}= (1-p)\frac{\mathbb{I}}{4}+p|\beta_{00}\rangle \langle\beta_{00}|,
\end{eqnarray}
where $p\in[0,1]$, $\mathbb{I}$ is the $4\times4$ identity matrix and the mnemonic notation $|\beta_{xy}\rangle$ can be understood via the equations
\begin{eqnarray}
|\beta_{xy}\rangle\equiv \frac{1}{\sqrt{2}}(|0,y\rangle+(-1)^x|1,\bar{y}\rangle)
\end{eqnarray}
in the standard two-qubit computational basis $\{|00\rangle,|01\rangle, |10\rangle, |11\rangle\}$. Here $\bar{y}$ is the negation of $y$ \cite{NC}. Here, the states $|0\rangle$ and $|1\rangle$
correspond to spin up and spin down states, respectively.
\begin{eqnarray*}
|0\rangle=\left( {\begin{array}{c}
1 \\
0
\end{array}}
\right),\quad
|1\rangle=\left( {\begin{array}{c}
0 \\
1
\end{array}}
\right).
\end{eqnarray*}

From here on under the action of Hamiltonians the EoF, $l_1$-norm of coherence and MID of the outputs $\sigma_{AB}^{(j)} (j=1,2)$ will be respectively denoted by
$E(\sigma_{AB}^{(j)})$, $C_{l_1}(\sigma_{AB}^{(j)})$ and $MID(\sigma_{AB}^{(j)})$.

\begin{figure}[!btp]
\centering
\includegraphics[width=15 cm]{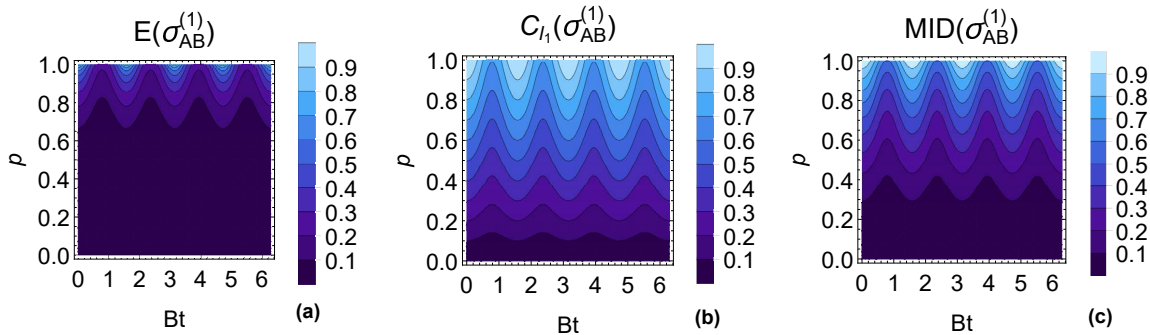}
\caption{(color online) Plots of (a) EoF, (b) $l_1$-norm of coherence and (c) MID versus state parameters $p$ and $Bt$ of the output $\sigma_{AB}^{(1)}$ under the action of Hamiltonian $H_1$
on the input state given by Eq. (23). For all plots, we take $\varphi=\pi/4$.}
\end{figure}

Now, we can calculate the EoF, $l_1$-norm of coherence and MID by the help of Eqs. (4), (6) and (9), respectively.
There measures of correlations for the output state $\sigma_{AB}^{(1)}$ under the action of Hamiltonian $H_1$ on the input state $\rho_{AB}^{(1)}$ is obtained as
\begin{eqnarray}
E(\sigma_{AB}^{(1)})&=&h\left(\frac{1+\sqrt{1-C\left(\sigma_{AB}^{(1)}\right)^2}}{2}\right),\\
C_{l_1}(\sigma_{AB}^{(1)})&=&\Big|p\sqrt{1-\cos^2\varphi \sin^2(2Bt)} \;\Big|,\\
MID(\sigma_{AB}^{(1)})&=& \frac{1}{4}[\left(1-p\right)\log\left(1-p\right)+\left(1+3p\right)\log\left(1+3p\right)-x_{+}\log x_{+}-x_{-}\log x_{-}]
\end{eqnarray}
where $x_{\pm}=1+p\pm 2p\cos\varphi \sin(2Bt)$ and concurrence is given by
\begin{eqnarray}
C\left(\sigma_{AB}^{(1)}\right)=C_{l_1}\left(\sigma_{AB}^{(1)}\right)-|1-p|/2.
\end{eqnarray}

We plot the evolutions of EoF, $l_1$-norm of coherence and MID as the function of $p$ and $Bt$ with $\varphi=\pi/4$ in Fig. 1. For all measures considered above the quantum correlations
(monotonically) increase with increasing values of initial state parameter $p$. However, $l_1$-norm of coherence is a more effective measure in determining the entanglement compared to
the other two measures. When quantum correlations are measured with $l_1$-norm of coherence entanglement approximately occurs at $p>0.1$ whereas, for the EoF and MID, this manifests itself at
$p>0.7$ and $p>0.3$, respectively. From Fig. 1(a). EoF attains the maximum at $p=1$ that naturally corresponds to a maximally entangled input state, that is, Bell state. Similar to EoF,
$l_1$-norm of coherence and MID oscillate with the increasing values of $Bt$ and they take place naturally take their maximum values for the maximally entangled inputs corresponding to $p=1$.
On the other hand, since the input $\rho_{AB}$ is a product state for $p=0$ the EoF, $l_1$-norm of coherence and MID vanish and there is no quantum correlation for these values of $p$.
Also, all correlation measures are independent of parameter $Bt$ for $\varphi=\pi/2$. Especially, EoF and $l_1$-norm of coherence are reduced to that of the values of the initial state, that is,
$E(\sigma_{AB}^{(1)})=h[(2+\sqrt{3(1-p)(1+3p)})/4]=E(\rho_{AB}^{(1)})$ and $C_{l_1}(\sigma_{AB}^{(1)})=p=C_{l_1}(\rho_{AB}^{(1)})$. On the other hand, MID can be calculated from Eq. (9).

As a second example, we consider a different two-qubit X-like state
\begin{eqnarray}
\rho_{AB}^{(2)}=p|\beta_{11}\rangle\langle \beta_{11}|
+\frac{1-p}{2}(|\beta_{01}\rangle\langle \beta_{01}|+|\beta_{00}\rangle\langle \beta_{00}|)
\end{eqnarray}
as an initial state in the standard basis. In this situation, we can again calculate the entanglement or correlation measures for the output state $\sigma_{AB}^{(2)}$ under the action of
Hamiltonian $H_1$ on the input $\rho_{AB}^{(2)}$ as follows
\begin{eqnarray}
E(\sigma_{AB}^{(2)})&=&h\left(\frac{1+\sqrt{1-C\left(\sigma_{AB}^{(2)}\right)^2}}{2}\right),\\
C_{l_1}(\sigma_{AB}^{(2)})&=&\frac{1}{2}\left(\Big|(1-p)\sqrt{1-\cos^2\varphi \sin^2(2Bt)}\;\Big|+|1-3p|\right),\\
MID(\sigma_{AB}^{(2)})&=& \frac{1}{4}[3\left(1-p\right)\log\left(1-p\right)+\left(1+3p\right)\log\left(1+3p\right)\nonumber\\
&&-2\left(1+p\right)\log\left(1+p\right)-y_{+}\log y_{+}-y_{-}\log y_{-}],
\end{eqnarray}
where $y_{\pm}=(1-p)[1\pm\cos\varphi \sin(2Bt)]$ and concurrence is given by
\begin{eqnarray}
C\left(\sigma_{AB}^{(2)}\right)=|1-3p|-C_{l_1}(\sigma_{AB}^{(2)}).
\end{eqnarray}

\begin{figure}[!btp]
\centering
\includegraphics[width=15cm]{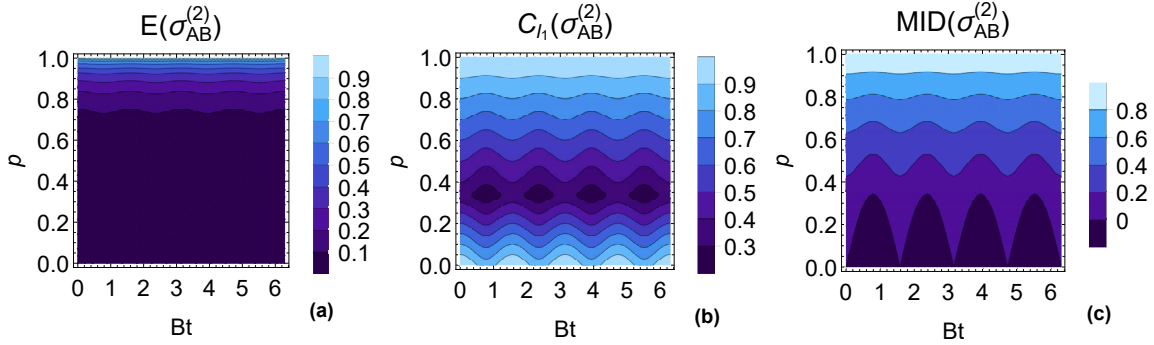}
\caption{(color online) Plots of (a) EoF, (b) $l_1$-norm of coherence and (c) MID versus the parameters $p$ and $Bt$ of the output $\sigma_{AB}^{(2)}$ under the action of Hamiltonian $H_1$
on the input state given by Eq. (29). For all plots, we take $\varphi=\pi/4$.}
\end{figure}

Fig. 2 displays the dynamics of the entanglement of the output $\sigma_{AB}^{(2)}$ under the action of the Hamiltonian $H_2$ on the input state given by Eq. (29) versus the parameters $p$ and $Bt$.
In Fig. 2(a), EoF has similar behavior to that of the previous case whereas differently from it, $l_1$-norm of coherence and MID show different behavior in Fig. 2(b) and (c). $l_1$-norm of coherence
attains a maximum in both the largest and the smallest value of $p$. For the former value, the initial state is reduced the maximally entangled while it corresponds to a sum of two Bell states with
equal probability for the latter value. On the other hand, for the certain values of the parameters, the phenomenon of entanglement sudden death (ESD) occurs in Fig. 2, especially (a) and (c),
and the entanglement revives after a while.

\subsection{Action of $H_2$ to Initial Werner-like States and Entanglement Dynamics}

In this case, for the output state $\sigma_{AB}^{(1)}$ under the action of Hamiltonian $H_2$ on the input $\rho_{AB}^{(1)}$, the concurrence is found to be $C\left(\sigma_{AB}^{(1)}\right)=(3p-1)/2$.
So, the correlation measures can be calculated as follows

\begin{eqnarray}
E(\sigma_{AB}^{(1)})&=&h\left(\frac{2+\sqrt{3(1+3p)(1-p)}}{4}\right),\\
C_{l_1}(\sigma_{AB}^{(1)})&=& p ,\\
MID(\sigma_{AB}^{(1)})&=& \frac{1}{4}[(1-p)\log(1-p)+(1+3p)\log(1+3p)\nonumber\\
&&-2(1+p)\log(1+p)].
\end{eqnarray}

It is noted that under this YBS the spectrum of the output system is invariably reduced to that of the input system no matter what the action of Hamiltonian $H_2$ is and the dynamical evolution of a two-qubit
system is clearly unaffected by YBS. So, the density matrix after the evolution is just the same up to a phase. So, from Eqs. (34)-(36) we say that all three measures depend only on the initial
state parameter $p$ and they naturally attain their maximum values at $p=1$ in which the input state corresponds to a maximally entangled state. The behavior of quantum correlations quantified by EoF,
$l_1$-norm of coherence and MID is depicted in Fig. 3 as a function of parameter $p$ with $\varphi=\pi/4$ for action of the $H_2$ on the initial state $\rho_{AB}^{(1)}$. Initially, EoF monotonically
decreases for $p<1/3$ and ESD occurs and then it increases with increasing values of $p$. On the other hand, we can find that quantum coherence is equal to $p$ and linearly changes with the parameter $p$
whereas MID monotonically increases with the parameter $p$. Parallel to the previous cases, $l_1$-norm of coherence is generally more effective than other measures to determine the dynamics of entanglement
except for small values of $p$.

\begin{figure}[!btp]
\centering
\includegraphics[width=8.5cm]{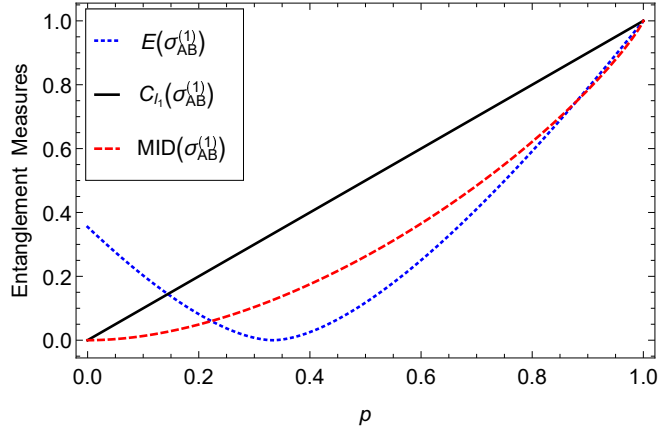}
\caption{(color online) Plots of entanglement or correlation measures versus the initial state parameter $p$ of the output $\sigma_{AB}^{(1)}$ under the action of Hamiltonian $H_2$
on the input state $\rho_{AB}^{(1)}$ given by Eq. (23).}
\end{figure}

\begin{figure}[!btp]
\centering
\includegraphics[width=15cm]{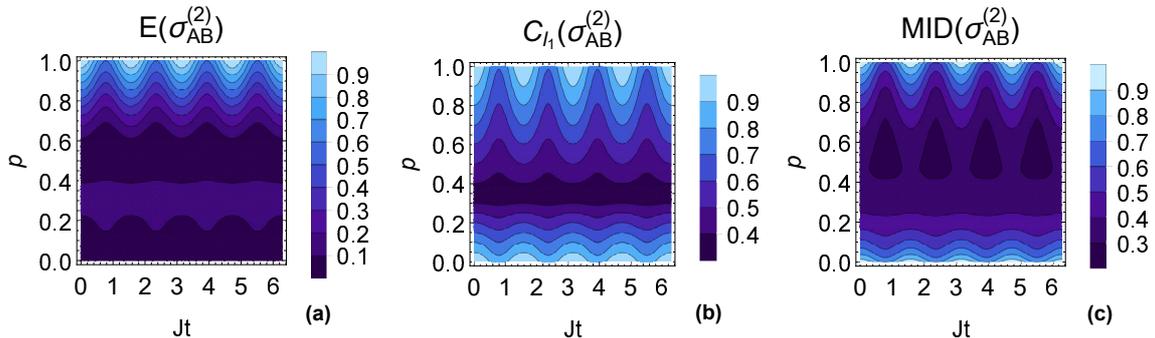}
\caption{(color online) Plots of (a) EoF, (b) $l_1$-norm of coherence and (c) MID versus the initial state parameter $p$ of the output $\sigma_{AB}^{(2)}$ under the action of Hamiltonian $H_2$
on the input state given by Eq. (29). For all plots, we take $\varphi=\pi/4$.}
\end{figure}

For the second initial state $\rho_{AB}^{(2)}$, the correlation measures can be found to be
\begin{eqnarray}
E(\sigma_{AB}^{(2)})&=&h\left(\frac{1+\sqrt{1-C\left(\sigma_{AB}^{(2)}\right)^2}}{2}\right),\\
C_{l_1}(\sigma_{AB}^{(2)})&=&\frac{1}{2}\left(|1-p|+\Big|(1-3p)\sqrt{1-\cos^2\varphi \sin^2(2Jt)}\;\Big|\right),\\
MID(\sigma_{AB}^{(2)})&=& -h(p)-\left(\frac{1-p}{2}\right)\log(1-p)-z_{+}\log z_{+}-z_{-}\log z_{-},
\end{eqnarray}
where $z_{\pm}=[1+p\pm(1-3p)\cos\varphi \sin(2Jt)]/4$ and concurrence is given by
\begin{eqnarray}
C\left(\sigma_{AB}^{(2)}\right)=C_{l_1}(\sigma_{AB}^{(2)})-|1-p|
\end{eqnarray}

Fig. 4 shows the evolutions of correlation measures with respect to $p$ and $Jt$ for $\varphi=\pi/4$. EoF takes place the maximum for $p=1$ that corresponds to the initial entangled state in Fig. 4 (a)
whereas the others attain their maximum values for both the largest and the smallest value of $p$ at certain values of $Jt$ in Fig. 4 (b) and (c). It is noted that for some intermediate values of $p$
EoF tends only to zero and entanglement vanishes or in other words ESD occurs. Among all correlation measures, coherence has a wide entanglement spectrum compared to the others, that is, entanglement
lives more for any fixed value of the parameters. On the other hand, for $\varphi=\pm\pi/2$ in Eqs. (37)-(39) all correlation measures depend only on the parameter $p$ in which $H_2$ corresponds to the
2-qubit anisotropic Heisenberg XXZ model under an inhomogeneous magnetic field. So far, all these observations imply that there is an unavoidable loss of correlations and hence information flows to the
environment. However, it can be concluded that the decrease of quantum correlations can be mitigated by adjusting the parameters and by choosing the appropriate initial states under the actions of the
different Hamiltonians constructed by YBE.

Finally, the third Hamiltonian $H_3$ corresponds to an anisotropic two-qubit Heisenberg XYZ chain for $\varphi=-\pi/2$. Here $B=\gamma J$ with the $\gamma$ is the anisotropy parameter.
Under the action of Hamiltonian $H_3$, since the correlation measures for outputs are the same as that of the first one it is easy to find that in the third YBS we get the same result
as the first one for all initial states.

\section{Concluding Remarks}
In this paper, some Hamiltonians have been constructed by the unitary YBMs $R_{i,i+1}(\theta, \varphi)$ from a Hamiltonian $H_0$ describing the nearest spin-spin interaction where parameters $\theta$
and $\varphi$ are time-independent. We have studied the behavior of quantum correlations quantifying the entanglement content of a quantum state for two-qubit systems under the actions of unitary Yang-Baxter
channels or the Yang-Baxterization approach. Our results clearly show that the actions of different YBMs have different effects on the robustness of quantum correlations. We have observed that for some ranges
of the parameters the behaviors of the EoF, coherence and MID depend on the choice of the input state and of YBMs or quantum channel. Among these measures, we conclude that the coherence is much more effective
than the others in the sense of the quantifying entanglement. Compared to the other two measures, when the coherence is used to measure the correlations it is observed that entanglement can be extracted more from the
state.

Besides all these, a complimentary remark is that the cohering and decohering power of any quantum channels can be explored to better understand that the quantum channel creates and destroys the coherence
of the input quantum states \cite{Mani}. The former quantifies the maximal amount of coherence that it creates when acting on a completely incoherent state. The latter is the maximum amount by which it reduces
the coherence of a maximally coherent state.

Another important result is that by adjusting the parameters and using the different Hamiltonians constructed by YBEs, it is possible to obtain the relative increments in quantum correlations and all measures
may also take high values to improve a better quantum information task. It would be attractive and significant to investigate if relative enhancement in generic and distinctive behaviors of other correlation
measures not considered here could be possible for the Yang-Baxterization approach and to study the dynamics of entanglement in presence of noisy environment in the sense of Markovianity and non-Markovianity
under the Hamiltonians constructed by YBE. Our studies on these issues are in progress.

\section*{Acknowledgements}
This work was supported in part by the Scientific and Technological
Research Council of Turkey (TUBITAK).



\end{document}